\documentclass[fleqn,twoside]{article}
\usepackage{espcrc2}
\usepackage{graphicx}
\usepackage[figuresright]{rotating}

\newcommand{\AmS}{{\protect\the\textfont2
  A\kern-.1667em\lower.5ex\hbox{M}\kern-.125emS}}

\title{Sterile neutrinos at future long baseline experiments}

\author{D. Meloni\address[MCSD]{Universit\'a degli Studi Roma Tre, Via della Vasca Navale 84, 00146 Roma \& INFN, sezione di RomaTre}%
        }
       
\begin{document}

\begin{abstract}
In this talk we review the current status of sterile neutrino searches and discuss the potential of future long baseline experiments to study their properties.
\vspace{1pc}
\end{abstract}

\maketitle

\section{Introduction}
Neutrino conversion has been confirmed in many experiments; today we know that there are
only three active flavours accompanying e, $\mu$ and $\tau$ leptons; this means that there is 
only space for neutrinos with no electroweak interactions. They are called 
{\it sterile neutrinos}.
Being electroweak singlets, sterile neutrinos show their effects 
in mixing with active neutrinos; in order to describe these extra mixings,
we need to introduce several more parameters to account for new mass differences and 
mixing angles. The simplest scenario we can imagine is that with only one more mass eigenstate 
in the game; 
this gives rise to the so-called $(3 + 1)$ and $(2 + 2)$ schemes: the former is a smooth modification 
of the standard three neutrino case in which the fourth mass eigenstate is well separate from the 3-$\nu$
subspace whereas in the latter a large mass squared difference separates two couples of almost degenerate 
neutrinos. It turns out that these models do not 
satisfactory describe all of neutrino data, including the LSND results
\cite{Maltoni02}, because of 
strong constrains from solar, atmospheric  and null-result short baseline experiments
\cite{Maltoni04}. In particular, 
in the (2+2) scheme the extra sterile state cannot be simultaneously decoupled
from both solar and atmospheric oscillation 
\cite{giunti}, whereas 
the (3+1) scheme suffers from a tension between LSND and null-result short baseline
disappearance experiments 
\cite{Maltoni01}.
This is even worse after the MiniBooNE results \cite{Maltoni:2007zf}, implying that 
the LSND anomaly cannot be explained using sterile neutrinos.
However, disregarding the LSND result,
the (3+1) is still is a perfectly viable extension of the Standard Model.

\section{Parametrizing the new mixing matrix}

In the case of Dirac neutrinos, an unitary 4 $\times$ 4 matrix has 6 mixing angles and 3 CP-violating phases;
we need to accomodate these new parameters.
Among the different parameterizations available, we choose the one useful for long
baseline experiments \cite{Maltoni:2007zf}:
\begin{eqnarray}
U_{LBL} &=& R_{34}(\theta_{34}) R_{24}(\theta_{24}) R_{23}(\theta_{23},\delta_3) 
R_{14}(\theta_{14})\nonumber \cr
&& R_{13}(\theta_{13},\delta_2) R_{12}(\theta_{12},\delta_1)  
\end{eqnarray}
The choice of phases is made in such a way that 
$\delta_1$ is in $R_{12}$ and it drops in the two-mass dominance regime, 
$\delta_2$ is in $R_{13}$ and  it reduces to the standard Dirac phase in the
three-family limit; the other phase can be put anywhere; we choose to insert it in $R_{23}$.
 With this parametrization we can analyze 
$\nu_e$ and $\nu_\mu$ disappearance data and conclude that both new angles 
$\theta_{14}$ and $\theta_{24}$ should be small, at the level of $10^\circ$ at 
$90\%$ CL \cite{Donini:2007yf}. $\theta_{34}$ is less constrained and can be as large as $35^\circ$
at $90\%$ CL.
 
\section{Probabilities in matter}

To understand the details of the channels with the greatest sensitivity to the
four-family neutrino schemes, it is useful to obtain simple analytical expressions
for the different channels in matter, expanded in some small parameters. Here we choose to 
expand up to quadratic order in $s_{13}$, $s_{14}$, $s_{24}$, $s^2_{34}$, where $s_{ij}=\sin \theta_{ij}$. We get:
\begin{eqnarray}
\label{prob}
P_{ee}& =& 1+ \mathcal O(s^2_{13})  \\
\nonumber
P_{e\mu} &\sim& P_{e\tau}\sim P_{es}\sim \mathcal O(s^2_{13})\\
\nonumber
P_{\mu\mu} &=& 1- \sin^2 2\theta_{23}\sin^2 \frac{\Delta m_{\rm atm}^2 L}{4E}- \nonumber \\
&& 2 (A_n L)s_{24}s_{34} 
\sin^2 2\theta_{23} \cos \delta_3\sin \frac{\Delta m_{\rm atm}^2 L}{2E}\nonumber \\
\nonumber  
P_{\mu\tau}&=&  \sin^2 2\theta_{23} (1-s^2_{34})    \sin^2 \frac{\Delta m_{\rm atm}^2 L}{4E}
+\nonumber \\ \nonumber
&& s_{24}s_{34}\left[\sin \delta_3+2(A_n L)\cos \delta_3\right] \sin \frac{\Delta m_{\rm atm}^2 L}{2E}.
\end{eqnarray}
where $A_n=\sqrt{2}\,G_F\,n_n /2$ and $n_n$ is the neutron density.
From these expressions 
it can be easily seen that the $\nu_e$ decouples within this approximation. 
We can thus
conclude that the "classic" Neutrino Factory channels, such as the
"golden" one $\nu_e \to \nu_\mu$ \cite{Cervera:2000kp} and "silver" one $\nu_e \to\nu_\tau$ 
\cite{Donini:2002rm},
are of limited interest to study sterile
neutrinos.
The most relevant oscillation channels in this context are
the appearance one $\nu_\mu\rightarrow\nu_\tau$ and
disappearance one $\nu_\mu \to \nu_\mu$. 

\section{(3+1) sterile neutrinos at OPERA}
\begin{figure}[h!]
\caption{\it Sensitivity limit at 99\% CL in the ($\theta_{13}$, $\theta_{14}$) plane (left) and in the 
($\theta_{24}$, $\theta_{34}$) plane
(right) from the combined analysis of present data and a null result of the OPERA experiment,
assuming 1, 2, 3, 5 and 10 times the nominal CNGS intensity of $4.5 \times 10^{19}$ pot/year. The coloured regions
show the present bounds at 90\% and 99\% CL. Other details in \cite{Donini:2007yf}.}
\label{opera}
\hspace{-0.8cm}
\includegraphics[width=20pc]{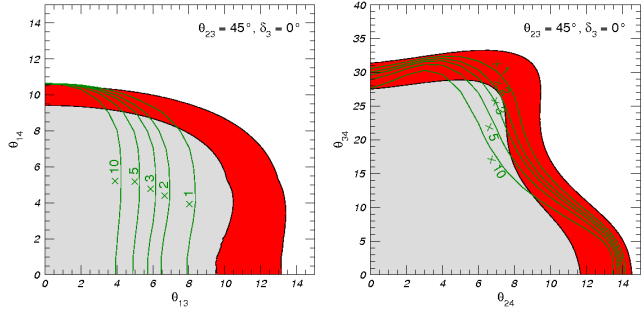}
\end{figure}

 In \cite{Donini:2007yf}, we have determined the presently allowed regions for all active-sterile mixing angles
and studied the OPERA capability to constrain them further using both the $\nu_\mu \to\nu_e$
and  $\nu_\mu \to\nu_\tau$ channels.
Our conclusions are the following: if the OPERA detector is exposed to the nominal
CNGS beam intensity, a null result (combined with present neutrino data) can improve a bit the present bound on $\theta_{13}$, but
not those on the active-sterile mixing angles, $\theta_{14}$, $\theta_{24}$ and $\theta_{34}$. 
If the beam intensity is
increased by a factor of 2 or beyond, not only the sensitivity to $\theta_{13}$ increases accordingly,
but a significant sensitivity to $\theta_{24}$ and $\theta_{34}$ is achievable (Fig.(\ref{opera})). On the other hand, the sensitivity on $\theta_{14}$ remains almost the same.

\section{Sterile Neutrinos at the Neutrino Factory}

\begin{figure}[h!]
\caption{\it Sensitivity to $\theta_{24}$ and $\theta_{34}$ of a Neutrino Factory at $L=3000$ Km. The other parameters are fixed to their best fit values \cite{Maltoni04}.}
\label{nufact}
\includegraphics[width=15pc]{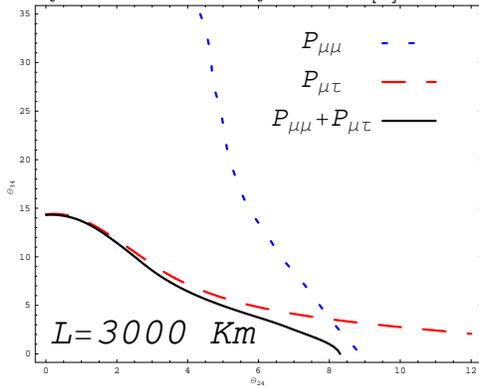} 
\end{figure}

Another interesting place where to look for sterile neutrinos is the Neutrino Factory. 
Looking at the probabilities $P_{\mu\mu}$
and $P_{\mu\tau}$ in Eq.(\ref{prob}), 
we  expect a reduced sensitivity on $\theta_{13}$ and  $\theta_{14}$ and  a larger sensitivity to  $\theta_{24}$ and $\theta_{34}$.
These considerations are confirmed by preliminary analysis in the context of a standard 50 GeV Neutrino Factory, 
with $2 \times 10^{20}$ muon decay/year and 4 years of data taking for both polarities.
The case of the baseline L = 3000 Km is shown in  Fig.(\ref{nufact}) \cite{wip}.
We look for muons with a Magnetized Iron Calorimeter assuming an efficiency for muons $\epsilon_\mu= 0.9$, and an overall systematic error at the level of $5\%$.
For taus, we use a Magnetized Emulsion Cloud Chamber assuming $\epsilon_\tau= 0.1$ and systematic error at the level of $10\%$. 
From the plot we can see that $\theta_{34}$ can be probed down to $\sim 15^\circ$ whereas the sensitivity on $\theta_{24}$ can be improved by a factor of 2 with respect to the current bounds.

\section{New CP phases  at the Neutrino Factory}

The formulae in Eq.(\ref{prob}) show that the main dependence on CP violation in the sterile sector
comes from the phase $\delta_3$.
In particular, CP asymmetry in the $P_{\mu\mu}$ channel are obviously due to matter effects
because $\delta_3$ appears under cosinus term; on the other hand, in the $\mu \to \tau$ channel, pure CP effects arise due to the $\sin \delta_3$ dependence. In  \cite{Dighe:2007uf} it has been shown that the CP asymmetries in the $\mu\mu$ and $\mu\tau$ channels have both a different behaviour (as a function of the neutrino energies) when evaluated for three or four neutrinos; thus these observables are an efficient
discriminator among the two models of neutrino oscillation. Obviously, when discussing CP violation, we also have to deal with the problem of degeneracies; this has been discussed in
\cite{Goswami:2008mi}, where the authors pointed out that, in the $\mu-\tau$ channel, a number of correlations and degeneracies can affect the determination of the CP phase $\delta_3$: 
the $sign[\Delta m^2_{23}] - \delta_3$ and the $\delta_3 - (\pi - \delta_3)$ degeneracies and the
$(s_{24} s_{34} - \delta_3)$ correlation. The problem can be mitigated using different combinations of detector masses and baselines for an OPERA-like and liquid argon (LAr) detectors. 
In particular, the allowed regions in the $(s_{24} s_{34} - \delta_3)$-plane can be strongly reduced using a 100 Kt LAr at $L=3000$ Km and a 0.1 Kt detector at $L=2$ Km. 

The Neutrino Factory can also be used to understand whether it is possible to confuse a (3 + 1) CP-conserving scenario with a 3-family CP-violating model. The answer to this question has been given in \cite{Donini:2001xp}, using the following algorithm applied to the golden channel: ``data'' are generated in the CP-conserving 
(3 + 1) scheme for some value of
$\theta_{14}$ and $\theta_{24}$ whereas $\theta_{13}$ and $\theta_{34}$  are left free; 
for any ($\theta_{13}$-$\theta_{34}$) we fit this data with the 3-family model;
if successful, the point ($\theta_{13}$-$\theta_{34}$) is a point in which confusion is possible.
The output of this analysis is that for $\theta_{14} \sim \theta_{24} > 5^\circ$, a Neutrino Factory at $L=732$ Km is very useful to tell 3 from 4 neutrino models; any combination of this setup with other baselines performs even better. In the points in which confusion is possible, the corresponding values of the standard $\delta$ are not too large: for $\sim$ 6000 successful fits, we get $-15^\circ < \delta < 15^\circ$ in the 31\%, 50\% and
37\% of the cases at the baselines $L=732,3500$ and $7300$ Km, respectively.

I wish to thank the organizers of NOW 2008,
Prof. Fogli and Prof. Lisi, for the very stimulating
atmosphere during the workshop.

\end{document}